\documentclass[showpacs,preprintnumbers,amsmath,amssymb]{revtex4} 
\usepackage{mathrsfs}
\usepackage{bm}
\usepackage{amsmath}

\newcommand{\be}{\begin{equation}}
\newcommand{\ee}{\end{equation}}
\newcommand{\bea}{\begin{eqnarray}}
\newcommand{\eea}{\end{eqnarray}}

\newcommand{\mrm}[1]{{\mathrm{#1}}}

\newcommand{\e}{\mathrm{e}}

\renewcommand{\vec}[1]{{\mathbf{#1}}}
\newcommand{\jvec}{\vec{j}}

\newcommand{\ompl}{\omega_{\mrm{pl}}}

\begin{document}
\title{A classical theory for second-harmonic generation from metallic nanoparticles}
\author{Yong Zeng$^{1}$\footnote{zengy@acms.arizona.edu}, Walter Hoyer$^{2}$\footnote{The first and second author contributes equally to this article.}, Jinjie Liu$^{1}$, Stephan W. Koch$^{2}$, Jerome V. Moloney$^{1}$}
\affiliation{1. Arizona Center for mathematical Sciences,
University of Arizona, Tucson, Arizona 85721\\2. Department of
Physics and Material Sciences Center, Philipps University, Renthof
5, D-35032 Marburg, Germany}
\input epsf
\begin{abstract}
In this article, we develop a \textit{classical} electrodynamic
theory to study the optical nonlinearities of metallic
nanoparticles. The quasi-free electrons inside the metal are
approximated as a classical Coulomb-interacting electron gas, and
their motion under the excitation of an external electromagnetic
field is described by the plasma equations. This theory is further
tailored to study second-harmonic generation. Through detailed
experiment-theory comparisons, we validate this classical theory
as well as the associated numerical algorithm. It is demonstrated
that our theory not only provides qualitative agreement with
experiments, it also reproduces the overall strength of the
experimentally observed second-harmonic signals.
\end{abstract}
 \maketitle

\section{Introduction}
Optical second-harmonic generation (SHG) from a metal (silver)
surface was first observed in 1965 \cite{brown}, four years after
the first observation of SHG from quartz in 1961 \cite{franken}.
In the following fifty years, a number of important features of
SHG from metallic surfaces have been founded such as (1) second
harmonic (SH) intensities can be enhanced more than an order of
magnitude by coupling incident light into surface polariton
resonances at metal surfaces \cite{sipe1}; (2) SHG from surfaces
of centrosymmetric metals is anisotropic, the strength of the SH
response thus depends on the relative orientation of the incident
field and the crystal axes \cite{guidotti,sipe2}; (3) Because of
local-field enhancement, SHG is very sensitive to surface
roughness and chemical processes such as adsorption and
electrochemical reactions \cite{chen1,chen2}. On the theoretical
side, different approaches on both phenomenological and
microscopic levels have been developed to analyze SH response from
metals \cite{janz}, such as classical Boltzman equation approach
\cite{jha}, hydrodynamic model
\cite{bloembergen,rudnick,sipe3,eguiluz,hua,maytorena},
phenomenological formalism in terms of the fundamental tensor
elements of the SH susceptibility
\cite{chen1,sionnest,sipe2,dadap,stockman,dadap2,li,cao,bachelier}
and the self-consistent density functional formalism (see
Ref.\cite{heinz,liebsch} and the cited references).

Recently, a renaissance of scientific interests appears in the
quadratic nonlinearities of metallic nanostructures and
nanoparticles (NPs) partially owing to the significant
localizations of electromagnetic (EM) field induced by the
plasmonic oscillations of the conduction electrons inside the
metal
\cite{john,maier,anceau,bouhelier,nahata,krause,nappa,neacsu,mcMahon,klein,shan,abe,nieuwstadt,kujala1,canfield,klein2,mcl,feth,kujala2,kim,maeda,husu,rossi}.
More specifically, SHG were experimentally observed from different
geometric configurations such as sharp metal tips
\cite{bouhelier,neacsu}, periodic nanostructured metal films
\cite{nahata}, imperfect spheres \cite{nappa,shan}, split-ring
resonators \cite{klein,klein2} and their complementary
counterparts \cite{feth}, metallodielectric multilayer
photonic-band-gap structures \cite{mcl}, T-shaped \cite{klein2}
and L-shaped NPs \cite{kujala1,kujala2}, noncentrosymmetric
T-shaped nanodimers \cite{canfield,husu} and ``fishnet" structures
\cite{kim}.

It should be emphasized that these subwavelength NPs and
one-dimensional interfaces have different structural symmetries,
and these differences further lead to significant consequences.
For ideally infinite interfaces, the dominant SH dipole source
appears only at the interface between centrosymmetric media where
the inversion symmetry is broken, although higher order multipole
sources provide a relatively small bulk SH polarization density.
The SH polarization density is thus significantly localized in a
\textit{surface} region a few Angstrom wide, and sensitively
influenced by the details of the surface electronic structure. On
the other hand, for low-symmetric or even asymmetric NPs, such as
gold split-ring resonators, SH dipolar polarizability may be
presented in the whole \textit{volume} and not limited at the
interface \cite{finazzi}. Consequently, the overall shape of the
NP plays a significant role in determining the SH response. To
analyze the quadratic nonlinearities of these metallic NPs, we
therefore propose that complicated microscopic models of the
interfaces are not required, and an easy-to-implement classical
model is sufficient.

The paper is organized as follows. Section 2 presents a classical
electrodynamic model which describes the nonlinearities induced by
Coulomb-interacting electron gas in metals. Using small
nonlinearity approximation, this classical model is further
tailored in Section 3 to treat second-order generation. Section 4
gives a detailed comparison between theoretical results and the
corresponding experiments. Discussion, conclusion and
acknowledgement are presented in Section 5 and Section 6,
respectively.

\section{Classical electrodynamic model}
In our model, the motion of quasi-free electrons inside a metal is
described classically. Quantum mechanical Coulomb correlations and
exchange contributions are thus missing while the classical
Coulomb interaction (i.e.\ the Hartree term in a quantum
mechanical derivation) is fully included \cite{walter}.
Furthermore, the Coulomb scattering due to higher-order quantum
corrections is phenomenologically described via an inverse decay
time $\gamma=1/\tau$. The electrons inside the metal are described
via their number density $n_{e}$ and velocity $\mathbf{u}_{e}$. In
our classical model, these quantities are continuous functions of
position $\mathbf{r}$ and time $t$
\cite{boyd,jeffrey,shen,robert}. We further assume the mass of
ions are infinite. Consequently, the ionic density
$n_0(\mathbf{r})$ is time-independent and only the electrons can
move and contribute to the current density. In other words, an
infinite barrier surface potential is assumed and
$n_0(\mathbf{r})$ is constant within the metal and zero outside
the metal.

We begin with two equations for electronic number density
$n_e(\mathbf{r},t)$ and the velocity field
$\mathbf{u}_e(\mathbf{r})$,
\begin{eqnarray}
&&\frac{\partial n_{e}}{\partial
t}+\nabla\cdot\left(n_{e}\mathbf{u}_{e}\right)=0,
\label{eq:ddt_ne}
\\
&& m_{e}\left(\frac{\partial}{\partial
t}+\mathbf{u}_{e}\cdot\nabla\right)\mathbf{u}_{e}=-e\left(\mathbf{E}+\mathbf{u}_{e}\times\mathbf{B}\right).
\label{eq:ddt_ue}
\end{eqnarray}
Here, the first equation is the usual continuity equation
expressed in terms of carrier density instead of charge density.
The second equation is the generalization of Newton's equation to
the case of a continuous field. The term in brackets on the
left-hand side is the so-called \textit{convective} or
\textit{material} derivative, which is a result of the description
of electrons in terms of a continuous density and can be formally
derived from the quantum mechanical Wigner distribution
\cite{walter}. More intuitively, it can be understood as the time
derivative of an electron taken with respect to a coordinate
system which is itself moving with velocity
$\mathbf{u}_{e}(\mathbf{r}(t),t)$, given by \cite{jeffrey}
\begin{equation}
\frac{d \mathbf{u}_{e}}{dt}=\frac{\partial
\mathbf{u}_{e}}{\partial t}+\left[\frac{d
\mathbf{r}(t)}{dt}\cdot\nabla\right]\mathbf{u}_{e}=\left(\frac{\partial}{\partial
t}+\mathbf{u}_{e}\cdot\nabla\right)\mathbf{u}_{e}.
\end{equation}

To describe the interaction between the classical electron gas and
the external EM fields self-consistently, we couple
Eqs.~(\ref{eq:ddt_ne}) and Eqs.~(\ref{eq:ddt_ue}) to Maxwell's
equations by defining the charge density and the current density
\begin{eqnarray}
\rho(\mathbf{r},t)&=&
e\left[n_0(\mathbf{r})-n_e(\mathbf{r},t)\right],
\label{eq:def_rho}
\\
\mathbf{j}(\mathbf{r},t)&=& -en_e(\mathbf{r},t)
\mathbf{u}_e(\mathbf{r},t)=\left[\rho(\mathbf{r},t)-e
n_0(\mathbf{r})\right]\mathbf{u}_e(\mathbf{r},t), \label{eq:def_j}
\end{eqnarray}
in terms of the electronic number density and velocity field.
Using these definitions and the equations of motion,
Eqs.~(\ref{eq:ddt_ne}) and Eqs.~(\ref{eq:ddt_ue}), we achieve
\begin{eqnarray}
\frac{\partial\rho}{\partial t} & = & -\nabla\cdot\mathbf{j},
\label{eq:ddt_rho}
\\
\frac{\partial \mathbf{j}}{\partial t} &=&
\sum_{k}\frac{\partial}{\partial r_{k}}\left(\frac{\mathbf{j}
j_{k}}{e n_0-\rho}\right)+\frac{e^2 n_0}{m_{e}}\mathbf{E}-
\frac{e}{m_{e}}\left[\rho\mathbf{E}+
\mathbf{j}\times\mathbf{B}\right]-\gamma\mathbf{j},
\label{eq:ddt_j}
\end{eqnarray}
where we have added a phenomenological term $-\gamma\mathbf{j}$ to
describe the current decay due to Coulomb scattering. The Lorentz
force describes a change in momentum due to an applied force. The
first term on the right-hand side, resulting from the convective
derivative, describes an increase or decrease of momentum simply
due to an accumulation or depletion of electrons at a certain
spatial point.

Equations~(\ref{eq:ddt_rho}) and~(\ref{eq:ddt_j}) have to be
coupled to Maxwell's equations. The final full set of equations to
be solved by a numerical scheme read as
\begin{eqnarray}
\frac{\partial\mathbf{B}}{\partial t} &=& -\nabla\times\mathbf{E},
\label{eq:ddtB}
\\
\frac{\partial\mathbf{E}}{\partial t} &=& c^2
\nabla\times\mathbf{B}-\frac{1}{\epsilon_0}\mathbf{j},
\label{eq:ddtE}
\\
\frac{\partial\mathbf{j}}{\partial t} &=&\frac{e^2
n_0}{m_{e}}\mathbf{E}-\gamma\mathbf{j}+\sum_{k}\frac{\partial}{\partial
r_{k}}\left(\frac{\mathbf{j}j_{k}}{e n_0-\rho}\right)-
\frac{e}{m_{e}}\left[\rho\mathbf{E}+\mathbf{j}\times\mathbf{B}
\right], \label{eq:ddtj}
\end{eqnarray}
where the charge density $\rho$ has to be viewed as a function of
the electric field since each occurrence of $\rho$ can be replaced
by the relation
\begin{equation}
\rho=\epsilon_0\nabla\cdot\mathbf{E}. \label{eq:rho}
\end{equation}
This set of equations couples the dynamics of the EM field to the
dynamics of the carriers described by their current density
$\mathbf{j}$. It should be mentioned that Equation~(\ref{eq:ddtj})
contains rich physics. The first two terms represent the linear
collective oscillation of the electrons with respect to the
background ionic density $n_0(\vec{r})$, and the last two terms
introduce three distinct sources for nonlinearities of the plasma.
The second and third source are the well-known electric- and
magnetic-component of the Lorentz force, respectively. The first
source term is a generalized divergence originating from the
convective time derivative of the electron velocity field
$\mathbf{u}_{e}$ mentioned above.

\section{Perturbative Expansion of Nonlinearities}

In order to obtain a simplified set of equations more suitable for
a numerical approach we expand every quantity in a power series of
the peak electric-field amplitude $|E_\mrm{exc}|$ of the
excitation pulse. Formally, we can write
\begin{eqnarray}
\mathbf{E}(\mathbf{r},t)& = &\sum_j\mathbf{E}^{(j)}(\mathbf{r},t),
\\
\mathbf{B}(\mathbf{r},t)& = &\sum_j\mathbf{B}^{(j)}(\mathbf{r},t),
\\
\mathbf{j}(\mathbf{r},t)& = &\sum_j\mathbf{j}^{(j)}(\mathbf{r},t),
\end{eqnarray}
where the functions $\mathbf{E}^{(j)}$, $\mathbf{B}^{(j)}$, and
$\mathbf{j}^{(j)}$ scale like $|E_\mrm{exc}|^j$. A similar
expansion automatically holds for the charge density by inserting
Eq.(12) into Eq.(11)
\begin{equation}
\rho(\mathbf{r},t)=\epsilon_{0}
\sum_j\nabla\cdot\mathbf{E}^{(j)}(\mathbf{r},t).
\end{equation}

Separating different orders, we obtain the linear response of the
metal via
\begin{eqnarray}
\frac{\partial\mathbf{B}^{(1)}}{\partial t} &=&
-\nabla\times\mathbf{E}^{(1)}, \label{eq:ddtB_lin}
\\
\frac{\partial\mathbf{E}^{(1)}}{\partial t} &=& c^2
\nabla\times\mathbf{B}^{(1)}-\frac{1}{\epsilon_0}
\mathbf{j}^{(1)}, \label{eq:ddtE_lin}
\\
\frac{\partial \mathbf{j}^{(1)}}{\partial t} &=&
-\gamma\mathbf{j}^{(1)} + \frac{e^2 n_0}{m_{e}} \mathbf{E}^{(1)}.
\label{eq:ddtj_lin}
\end{eqnarray}
This is equivalent to the well-known Drude model of a metal whose
bulk plasma frequency $\omega_{p}^{2}=e^2 n_0/m_{e}\epsilon_{0}$,
as can be easily seen by Fourier transformation \cite{born}.

The second-order fields describe the lowest-order nonlinearity of
the metal and are given by
\begin{eqnarray}
\frac{\partial\mathbf{B}^{(2)}}{\partial t} &=&
-\nabla\times\mathbf{E}^{(2)}, \label{eq:ddtB_SH}
\\
\frac{\partial\mathbf{E}^{(2)}}{\partial t} &=& c^2
\nabla\times\mathbf{B}^{(2)} - \frac{1}{\epsilon_0}
\mathbf{j}^{(2)}, \label{eq:ddtE_SH}
\\
\frac{\partial \mathbf{j}^{(2)}}{\partial t} &=&
-\gamma\mathbf{j}^{(2)} + \frac{e^2 n_0}{m_{e}} \mathbf{E}^{(2)}
+\mathbf{S}^{(2)}, \label{eq:ddtj_SH}
\end{eqnarray}
together with the nonlinear source term
\begin{equation}
\mathbf{S}^{(2)}=\sum_{k}\frac{\partial}{\partial r_{k}}
\left(\frac{\mathbf{j}^{(1)} j^{(1)}_{k}}{e n_0}\right)-
\frac{e}{m_{e}}\left[\epsilon_0 \left( \nabla\cdot
\mathbf{E}^{(1)}\right)\mathbf{E}^{(1)}+
\mathbf{j}^{(1)}\times\mathbf{B}^{(1)}\right], \label{eq:S2}
\end{equation}
where $k$ represents $x$, $y$ and $z$ coordinate. The homogeneous
part of this set of equations is identical to the first-order
equations such that the propagation of the SH field is modified by
the Drude response of the metal. The source term is expressed
fully in terms of the first-order fields such that the sets of
Eqs.~(\ref{eq:ddtB_lin})--(\ref{eq:ddtj_lin})
and~(\ref{eq:ddtB_SH})--(\ref{eq:S2}) can be solved separately.

We want to stress that no approximations have been done yet except
the expansion in orders of the exciting electric field. All fields
are real quantities and no expansion in terms of ``phase factor
times slowly varying envelop'' (see the Appendix) has been done so
far. In principle, these equations can be numerically solved, and
a switch-off analysis can be further utilized to distinguish the
contribution of three nonlinear sources.

A three-dimensional finite-difference time-domain (FDTD) algorithm
is employed to numerically solve the first-order and second-order
equations separately \cite{taflove}. Yee's discretization scheme
is employed so that all field variables are defined in a cubic
grid. Electric and magnetic fields are temporally separated by
half a time step, they are also spatially interlaced by half a
grid cell. Central differences in both space and time are then
applied to Eqs.~(\ref{eq:ddtB_lin})--(\ref{eq:S2}) \cite{yee}. In
addition, all the NPs studied are made of gold with Drude-type
permittivity approximated as
\begin{equation}
\epsilon(\omega)=1.0-\frac{\omega_{p}^{2}}{\omega(\omega+i\gamma)}
\end{equation}
with the bulk plasma frequency $\omega_{p}=1.367\times 10^{16}
s^{-1}$ and the phenomenological collision frequency
$\gamma=6.478\times 10^{13} s^{-1}$ \cite{klein,feth,palik}
(please notice that we employ $\omega_{p}$, not $n_0$, in the
simulations). In order to describe the energy conversion
efficiency in the nonlinear-optical process, we define a
normalized SH intensity,
\begin{equation}
\eta=|\mathbf{E}^{(2)}(2\omega_{0})/\mathbf{E}^{(1)}(\omega_{0})|^{2},
\end{equation}
to measure the strength of the far-field SH signal, where
$\omega_{0}$ is the frequency of the incident
fundamental-frequency (FF) wave.

\begin{figure}
\epsfxsize=350pt \epsfbox{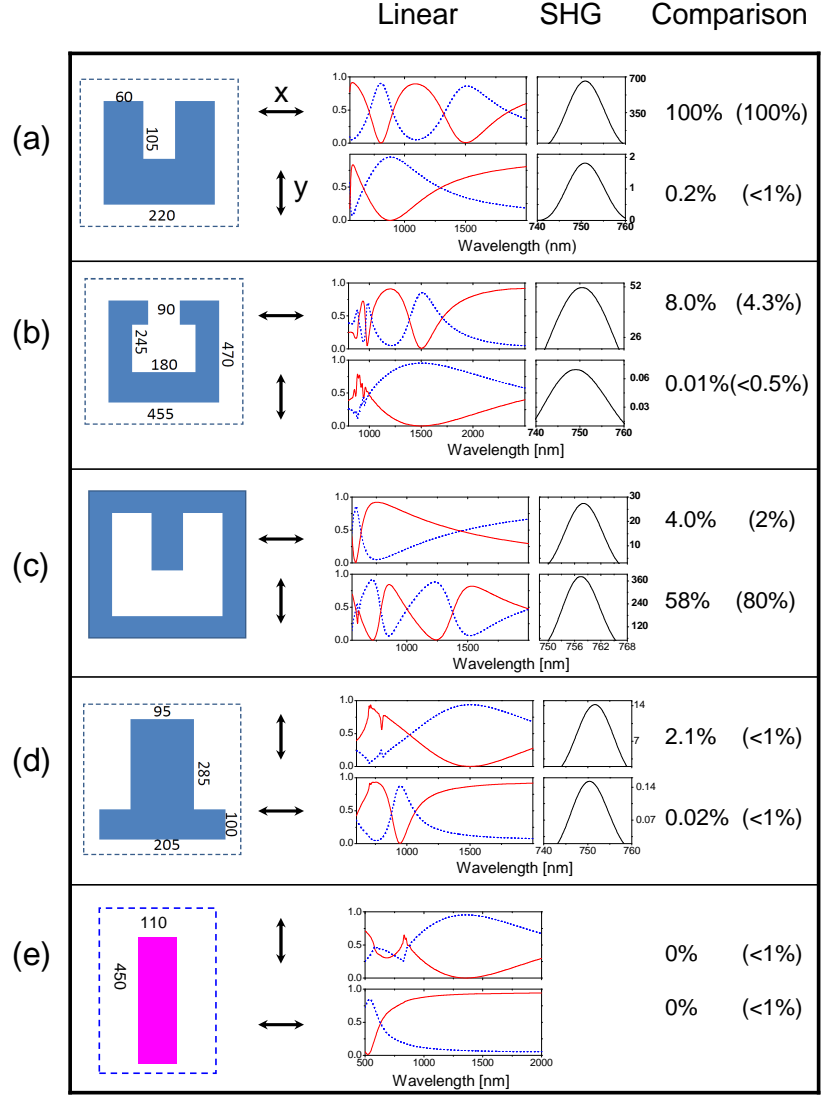} \vspace*{-0.5cm}
\caption{(Color online) Comparisons of numerical simulations and
experiments for different arrays of gold nanoparticles. The
different columns (from left to right) show shape of
nanoparticles, the polarization of the incident light (indicated
by the arrows), the linear transmission (solid) and reflection
(dashed) spectra, theoretical SHG spectra (amplified 13 orders of
magnitude), the relative strengths obtained by the theory and the
corresponding experiments (inside brackets). For all the
structures, the polarization of the generated second-harmonic
waves are along the $y$ direction. The illuminating
fundamental-frequency wave has wavelength around 1500 nm and
amplitude of $2\times 10^{7}$ (V/m). (a) The U-shape particle
corresponds to the experimental sample shown in Fig.(1A) of
Ref.\cite{klein} (also at Fig.(1a) of Ref.\cite{klein2}), with
lattice constant $a_{x}=a_{y}=$305 nm. (b) The C-shape particle
corresponds to the experimental sample shown in Fig.(1C) of
Ref.\cite{klein} (also at Fig.(1c) of Ref.\cite{klein2}), with
$a_{x}=567.5$ nm and $a_{y}=590$ nm. (c) The inverse-U-shape
particle corresponds to the experimental sample shown in
Ref.\cite{feth}, with $a_{x}=a_{y}=$305 nm. (d) The T-shape
particle corresponds to the experimental sample shown in Fig.(2c)
of Ref.\cite{klein2}, with $a_{x}=295$ nm and $a_{y}=465$ nm. (e)
The rectangle-shape corresponds to the experimental sample shown
in Fig.(2b) of Ref.\cite{klein2}. Here $a_{x}$ and $a_{y}$ are the
lattice constants along $x$ and $y$ direction, respectively. All
nanoparticles dimensions shown are in nanometer.}
\end{figure}

Our interests in the present article are limited to arrays of
metallic NPs (also named as planar metamaterial
\cite{klein,klein2,kim}) with normal incidence, the computational
domain is therefore arranged as follows: an array of NPs is placed
in the middle of the space with its top and bottom surfaces
positioned perpendicular to the $z$ direction; plane waves
propagating along the $z$ axis are generated by a total
field/scattering field technique \cite{taflove}; perfect matched
absorbing boundary conditions are applied at the top and bottom of
the computational space together with periodic boundary conditions
on other boundaries \cite{berenger}; the structure studied extends
periodically in the $x$ and $y$ directions, and only single unit
cell is needed in the computational space. In addition, in all the
following simulations, the size of the spatial grid cell is fixed
as 2.5 nm, and the associate time step is 4.17 attosecond.

\section{Theory-Experiment comparison}

In this section detailed comparisons between the numerical
evaluations of our theory with the experiments done in two
independent laboratories are presented
\cite{klein,klein2,feth,canfield}. First, we consider a series of
experiments reported in Refs.\cite{klein,klein2,feth}, in which
NPs with different geometrical configurations are studied (see
Fig. 1). Among them the U-shaped NPs (split-ring resonators) are
known to possess negative effective permeabilities in certain
frequency regions and are generally referred to as magnetic
metamaterials \cite{klein,rockstuhl,lazarides,smith,sa}. Strictly
speaking, those ``metamaterials" are only the first step towards a
true three-dimensional bulk material. So far, most of the
metamaterials are rather two-dimensional arrays of unit cells to
study the fundamental properties of the NPs. These samples are
supported by infinite-thickness glass (with $\epsilon=2.25$)
substrate coated with a thin film of indium-tin-oxide (with
$\epsilon=3.8$), and the thicknesses of the gold and
indium-tin-oxide layers are 25 nm and 5 nm, respectively
\cite{klein,klein2,feth}. It should be emphasized that the
geometrical parameters of these NPs are chosen such that each
structure has a resonance around 1500-nm wavelength.

Our theoretical results are summarized in Fig. 1. We note that the
simulations qualitatively agree with the corresponding
experimental measurements. The SH signal strength emitted from the
U-shape particles with $x$-polarized FF incidence is found as
$6.6\times 10^{-11}$ from the simulation, which is quite close to
the experimental result of $2.0\times 10^{-11}$ \cite{klein2}
(Please notice that the SH strength reported in Ref.\cite{klein2}
has been corrected in the sequent erratum). Our simulation thus
reproduces the strength of the experimental SH signal. The
following important conclusions can be further extracted from Fig.
1:

(1) The polarization state of the far-field SH signal is always
y-polarized (Fig. 2) for both x- or y-polarized incident fields
(except the rectangle-shaped NPs). There thus exists a universal
selection rule, that is, a mirror symmetry of the metallic NPs in
one direction completely prohibits a polarization component of SHG
in that direction. This symmetry dependence can be explained as
follows. Within the electric dipole approximation, the far-field
SH electric field can be related to the incident FF field such
that \cite{boyd,shen}
\begin{equation}
\mathbf{E}(2\omega)=\overleftrightarrow{\chi}^{(2)}\cdot\mathbf{E}(\omega)\mathbf{E}(\omega),
\end{equation}
where $\overleftrightarrow{\chi}^{(2)}$ stands for a generalized
dyadic second-order nonlinear susceptibility. The $x$-coordinate
mirror symmetry results in the following vanishing elements,
$\chi^{(2)}_{xxx}$, $\chi^{(2)}_{xyy}$, $\chi^{(2)}_{yxy}$ and
$\chi^{(2)}_{yyx}$. Further representing the FF field as $
\mathbf{E}(z,\omega)=E_{0}e^{i(\omega t-k
z)}\left[\cos\theta\mathbf{e}_{x}+\sin\theta\mathbf{e}_{y}\right]$,
with $k=\omega/c$ being the wave number, $\theta$ being the
polarization angle and $\mathbf{e}_{x}$ ($\mathbf{e}_{y}$) being
the unit vector along the $x$ ($y$) direction, we obtain
\begin{eqnarray}
&&E_{x}(2\omega)=E_{0}^{2}\chi^{(2)}_{xxy}\sin2\theta,\cr &&
E_{y}(2\omega)=E_{0}^{2}\left[\chi^{(2)}_{yxx}\cos^{2}\theta+\chi^{(2)}_{yyy}\sin^{2}\theta\right].
\end{eqnarray}
The $E_{x}(2\omega)$ component is simply proportional to
$\sin2\theta$, and therefore vanishes for $\theta=0$ or
$\theta=\pi/2$, corresponding to $x$- or $y$-polarized FF
incidence. On the other hand, the relationship between
$E_{y}(2\omega)$ and $\theta$ is nontrivial, and it survives for
$\theta=0$ and $\theta=\pi/2$.

\begin{figure}
\epsfxsize=260pt \epsfbox{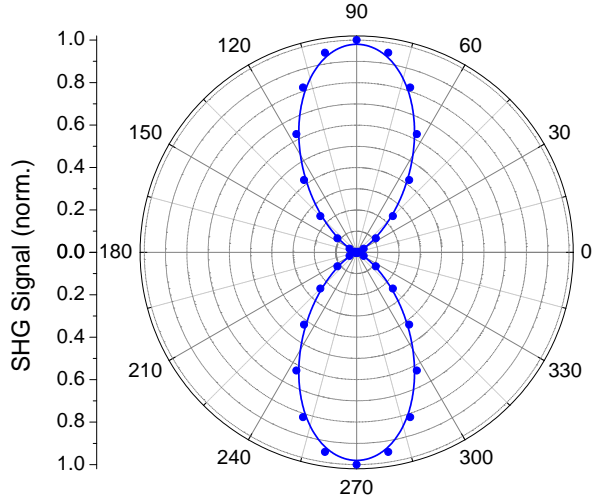} \vspace*{-5.5cm}
\caption{(Color online) The polarization state of the second
harmonic emission from an array of U-shape particles illuminated
with a $x$-polarized plane wave at the fundamental frequency. The
second-harmonic signal is a function of the measuring angle (not
the polarization of the incident field) $\theta$ ($\theta=0$
corresponds to the $x$ direction). The corresponding experiment
measurement is shown in Fig.(2B) of Ref.\cite{klein}.}
\end{figure}

(2) Similar to the fact that SHG at metal surfaces can be
significantly enhanced by coupling incident light into surface
polariton resonances \cite{janz}, the presence of structural
plasmonic resonances also can greatly enhance SHG from metallic
NPs. Moreover, different-order plasmonic resonances make different
contributions to the SHG. For example, an enlarged version of the
U-shape NP from Fig.(1a) possesses a second-order resonance
coincident with the fundamental resonance of the original. The SHG
emitting from the fundamental resonance of the original U is
considerably stronger than the SHG from the second-order resonance
of the larger structure, even without the perfect phase matching
requirement, because the near field enhancement is maximized for
the fundamental resonance \cite{john,robert,yong1,yong2}.

(3) It is found in the simulation that no SH signal is emitted
from the rectangle-shaped NPs in the far field. As stated earlier,
this is a direct consequence of the fact that an individual
rectangle-shaped NP possesses mirror symmetries along both $x$ and
$y$ directions. A dipole SH source is therefore forbidden and only
quadrupole sources (and higher-order multipoles) are allowed (the
retardation effects are negligible here since the thicknesses of
NPs are much smaller than the SH wavelength, while these effects
are found to excite nonlocal SH dipole for large-size gold
nanospheres \cite{nappa}). For a periodic array of
rectangle-shaped NPs with translational symmetry, the SH signal
from each individual NP interfere destructively. Therefore, only
near-field SH signal exists for the array. However, slight
far-field SH signal is observed in the experiment. This deviation
is believed to originate from the fact that the samples are not
rigorously inversion-symmetry because of the fabrication
imperfections (see the scanning electron micrograph shown in
Ref.\cite{klein2}).

Next we study the effect of the gap on the far-field SH strength
in noncentrosymmetric T-shaped gold nanodimers (Fig. 3). The
corresponding experiment is reported in Ref.\cite{canfield}, and
the scanning electron micrograph images of two dimers are shown in
Fig. 3. Obviously, these T-shaped dimers does not possess any
mirror symmetry along either $x$ or $y$ direction. The gold array
is 20 nm thick, with a lattice spacing of 400 nm. It is further
covered with a 20-nm protective layer of glass and supported with
an infinite-thickness glass substrate.

To include the relative difference between the configurations of
the samples, the NPs (for all gaps) employed in our simulations
are obtained by directly scanning the experimental samples, and
the computational cell consists of two T-shaped dimers. Our
numerical results are plotted in Fig. 3 and reproduce the
experimental observations qualitatively. More specifically, the
SHG for two configurations, YXX ($y$-polarized SH signal with
$x$-polarized fundamental fields) and YYY ($y$-polarized SH signal
with $y$-polarized fundamental fields), strongly depend on the
size of the gap, in a non-monotonically decreasing fashion. The
40-nm gap yields weak SHG responses for both configurations and
the largest SHG response occurs for YXX from the 2-nm gap. For
gaps smaller than 15 nm, the YYY response is much weaker than the
YXX response. On the other hand in a simulation of ideal
structures, without the geometrical variation of the dimers
induced by the fabrication imperfections, the YXX response
decreases monotonically with increasing gap. The non-monotonic
responses observed in the experiments thus arise from two sources,
the near-field enhancement around the NPs decreases with
increasing the gap, and variations of the overall shape of
different-gap nanodimers due to the fabrication imperfections.

\begin{figure}
\epsfxsize=260pt \epsfbox{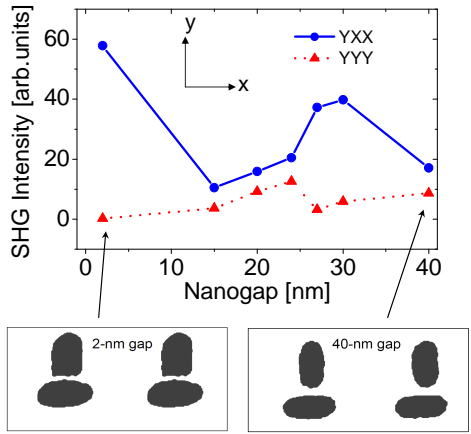} \vspace*{-5.5cm}
\caption{(Color online) Second harmonic generation in
noncentrosymmetric nanodimers with varying gaps. The configuration
such as YXX indicates $y$-polarized second-harmonic signal with
$x$-polarized fundamental field. The two images show the unit
cells as simulated, containing two structures which were digitized
from the scanning electron micrograph images of the experimental
samples (Ref.\cite{canfield}). The corresponding experimental
results are reported in Ref.\cite{canfield} (Fig.1 and Fig.3).}
\end{figure}

\section {Discussion and Conclusions}
The experiment-theory agreement presented above suggests that, in
contrast to an ideally infinite interface whose SH response
strongly localizes in its surface region, the overall shape of the
NP plays an important or even dominant role in determining
efficient SH emission \cite{finazzi}. Hence, even without an
accurate description of the surface electrons, our classical model
can not only successfully reproduce the experimental observations
qualitatively, but also reproduce the SH intensities.

It should be mentioned that Schaich developed an approach quite
similar to ours to study SHG by periodically-structured metal
surfaces \cite{schaich}. The major difference is that analytical
parametrizations of SHG at the (top) interface of a thick metal
slab \cite{corvi,liebsch,schaich2,liebsch2} are taken by assuming
the parametrization scheme still works even when the flat metal
surfaces are not of infinite extent but have edges and corners.
However, the validation of this assumption is unclear, especially
for subwavelength objects where the separation between two
neighboring edges may only be tens of nanometers and rapid
variations of the parametrizations are therefore expected.
Furthermore, this parametrization scheme limits the application of
his approach for NP with complicated boundaries such as the
nanodimers studied above as well as NPs with thin thicknesses. In
addition, there are some generalized theoretical works regarding
nonlinear metamaterials \cite{liu,feng} and the nonlinear
properties of negative-index metamaterials \cite{zharov,scalora}.

We want to point out that our classical model contains only the
influence of the conduction electrons and neglects contributions
from other sources such as core electrons and lattice phonons. It
therefore, for example, cannot correctly describe third-order
nonlinearities where the electronic polarization is negligible
comparing with other effects such as saturated atomic absorption
\cite{shen}. To include the third-order nonlinearities, we need to
add another current term
\begin{equation}
\mathbf{j}_{\rm
c}(t)=\chi^{(3)}\frac{d}{dt}|\mathbf{E}|^{2}\mathbf{E},
\end{equation}
where $\chi^{(3)}$ is the third-order nonlinear susceptibility,
which equals $7.56\times 10^{-19} \rm{m^{2}/V^{2}}$ for gold
\cite{robert}. The new set of equations has been utilized to study
third-harmonic generation from the NPs reported in
Ref.\cite{klein2}. It was demonstrated (see Appendix II) that our
simulations not only reproduce the overall strength of the
experimentally observed third harmonic signals, but also
qualitatively reproduce the structure dependent changes. As
expected, we found the third harmonic strength to be closely
related to the localization degree of the FF field inside the
metal.

In conclusion, a classical theory of second-harmonic generation
from metallic nanoparticles is presented. The conductor-band
electrons inside the metal are approximated as a classical
continuous plasmonic fluid, and its dynamics under an external
electromagnetic field are described by the plasma wave equations
self-consistently. A three-dimensional finite-difference
time-domain approach is further applied to solve these equations
numerically. By comparing theoretical results directly with the
corresponding experiments, it is demonstrated that our classical
theory, even without an accurate treatment of the surface
electrons, qualitatively captures the dominant physical mechanisms
of second-harmonic generations from metallic nanoparticles. This
agreement suggests that the second-harmonic emission from
nanoparticles depends strongly on their overall configurations.

\section {Acknowledgment}
The authors are grateful to Prof. Martti Kauranen at the Tampere
University for providing the SEM images of the experimental
samples. They also want to thank Prof. Martin Wegener and his
group at the Universit\"{a}t Karlsruhe, Dr. Jens F\"{o}rstner at
the Paderborn University, Prof. Moysey Brio, Dr. Miroslav Kolesik
and Dr. Colm Dineen at the University of Arizona for their
invaluable discussions. This work is supported by the Air Force
Office of Scientific Research (AFOSR), under Grant No.
FA9550-07-1-0010 and FA9550-04-1-0213. J. V. Moloney acknowledges
support from the Alexander von Humboldt foundation.

\section{Appendix I: approximation for quasi-monochromatic excitation}
For a quasi-monochromatic pulse with central angular frequency
$\omega_0$, one can classify the different contributions in terms
of their complex phase factor. For example, the linear electric
field is given by
\begin{equation}
\mathbf{E}^{(1)}(\mathbf{r},t)=\left[\widetilde{\mathbf{E}}^{(1)}(\mathbf{r},t)
\e^{-i\omega_0 t}+\mrm{c.c.}\right]\label{eq:linfreq}
\end{equation}
with the slowly varying complex field
$\widetilde{\mathbf{E}}^{(1)}$, while the second order field
\be
\vec{E}^{(2)}(\vec{r},t) = \widetilde{\vec{E}}^{(2)}_0(\vec{r},t)
+ \left[ \widetilde{\vec{E}}^{(2)}_2(\vec{r},t) \e^{-i 2 \omega_0
t} + \mrm{c.c.} \right] \label{eq:SHfreq} \ee has a
second-harmonic contribution proportional to the phase factor
$\e^{-i 2 \omega_0 t}$ multiplied with the slowly varying complex
amplitude $\widetilde{\vec{E}}^{(2)}_2$, as well as a slowly
varying low-frequency part $\widetilde{\vec{E}}^{(2)}_0$. The
magnetic field and the current can be expanded in a similar way.

As a next step, we want to express the source term from
Eq.~(\ref{eq:S2}) solely in terms of the linear electric field. To
that aim, we use the linear Eqs.~(\ref{eq:ddtB_lin})
and~(\ref{eq:ddtj_lin}) with the quasi-monochromatic approximation
of Eq.~(\ref{eq:linfreq}) and obtain
 \bea
 i \omega_0 \widetilde{\mathbf{B}}^{(1)} = \nabla\times\widetilde{\mathbf{E}}^{(1)}
\quad &\Longrightarrow& \quad \widetilde{\mathbf{B}}^{(1)} =
-\frac{i}{\omega_0} \nabla\times\widetilde{\mathbf{E}}^{(1)}\,,
\label{eq:B_lin}
\\
-i \omega_0 \widetilde{\jvec}^{(1)} =
-\gamma\widetilde{\jvec}^{(1)} + \frac{e^2 n_0}{m_{e}}
\widetilde{\vec{E}}^{(1)} \quad &\Longrightarrow& \quad
\widetilde{\jvec}^{(1)} = \frac{i}{\omega_0 + i \gamma} \frac{e^2
n_0}{m_{e}} \widetilde{\vec{E}}^{(1)}\,, \label{eq:j_lin}
\eea
where we have matched the terms with equal phase factor $\e^{- i
\omega_0 t}$.

Since every contribution to $\vec{S}^{(2)}$ in Eq.~(\ref{eq:S2})
is of the form of a product $A^{(1)} B^{(1)}$ between two first
order terms, these products according Eq.~(\ref{eq:linfreq}) can
be expressed as
\bea A^{(1)} B^{(1)} & = & \left(
\widetilde{A}^{(1)} \exp^{-i\omega_0 t} + \mrm{c.c.} \right)
\left( \widetilde{B}^{(1)} \exp^{-i\omega_0 t} + \mrm{c.c.}
\right) \nonumber
\\
& = & \left[ \widetilde{A}^{(1)} \widetilde{B}^{(1)} \exp^{- i 2
\omega_0 t} + \mrm{c.c.} \right] + \left[ \widetilde{A}^{(1)}
(\widetilde{B}^{(1)})^* + \mrm{c.c.} \right]\,.
\label{eq:compprod} \eea Thus, for the SH source
$\widetilde{\vec{S}}^{(2)}_{2}$, only the products of the slowly
varying complex fields have to be calculated. They can be computed
term by term and in the limit $\gamma = 0$ the first contribution
from the convective term is given by \be
\widetilde{\vec{S}}^{(2)}_{2} \Bigr|_{\mrm{conv}} =
\sum_{k}\frac{\partial}{\partial r_{k}}
\frac{\widetilde{\jvec}_{1} \widetilde{j}_{1,k}}{en_{0}} = -
\frac{e}{m_e} \frac{\epsilon_0}{\omega_0^2} \left[ ( \ompl^2
\widetilde{\vec{E}}^{(1)}\cdot \nabla) \widetilde{\vec{E}}^{(1)} +
\widetilde{\vec{E}}^{(1)} \left(\nabla \cdot  (\ompl^2
\widetilde{\vec{E}}^{(1)})\right) \right]\,, \label{eq:Sf_1} \ee
where the plasma frequency is defined as $\ompl^2(\vec{r}) = e^2
n_0(\vec{r})/(m_e \epsilon_0)$. The second term of
Eq.~(\ref{eq:S2}) -- the electric Lorentz force -- is already
expressed solely in terms of the electric field and the third,
magnetic term can be written as \be \widetilde{\vec{S}}^{(2)}_{2}
\Bigr|_{\mrm{magn}} = -\frac{e}{m_e}
\widetilde{\jvec}^{(1)}\times\widetilde{\vec{B}}^{(1)} =
\frac{e}{m_e} \frac{\epsilon_0}{\omega_0^2} \ompl^2 \, \left[
(\widetilde{\vec{E}}^{(1)} \cdot \nabla )
\widetilde{\vec{E}}^{(1)} - \frac{1}{2} \nabla \left(
\widetilde{\vec{E}}^{(1)} \cdot  \widetilde{\vec{E}}^{(1)} \right)
\right]. \ee Adding up all three contributions to the complex SH
source term $\widetilde{\vec{S}}^{(2)}_2$ is then given by \be
\widetilde{\vec{S}}^{(2)}_2 = - \frac{e}{m_e}
\frac{\epsilon_0}{\omega_0^2} \left[ \widetilde{\vec{E}}^{(1)}
\left(\nabla \cdot  (\ompl^2  \widetilde{\vec{E}}^{(1)})\right) +
\omega_0^2 \widetilde{\vec{E}}^{(1)} \left( \nabla\cdot
\widetilde{\vec{E}}^{(1)}\right) + \frac{\ompl^2}{2} \nabla \left(
\widetilde{\vec{E}}^{(1)} \cdot  \widetilde{\vec{E}}^{(1)} \right)
\right]. \ee Furthermore, from the first order wave equation, we
find that \be
\nabla\cdot\widetilde{\vec{E}}^{(1)}=\frac{1}{\omega_0^{2}}\nabla\cdot(\ompl^2
\widetilde{\vec{E}}^{(1)})\,, \label{eq:divE} \ee such that the SH
source can be accordingly simplified to \be
\widetilde{\vec{S}}^{(2)}_2 = - \frac{e\,\epsilon_0}{m_e} \left[ 2
\widetilde{\vec{E}}^{(1)} \left(\nabla \cdot
\widetilde{\vec{E}}^{(1)}\right) +
\frac{1}{2}\frac{\ompl^2}{\omega_0^2} \nabla \left(
\widetilde{\vec{E}}^{(1)} \cdot  \widetilde{\vec{E}}^{(1)} \right)
\right]. \ee

In a similar fashion, also the low-frequency source
$\widetilde{\vec{S}}^{(2)}_0$ can be derived. Repeating analogous
steps for the second term of Eq.~(\ref{eq:compprod}) we obtain \be
\widetilde{\vec{S}}^{(2)}_0 = \frac{e\,\epsilon_0}{m_e}
\frac{\ompl^2}{\omega_0^2} \nabla |\widetilde{\vec{E}}^{(1)}|^2\,,
\ee which is the well-known ponderomotive force.

In order to insert the nonlinear source into the differential
equation for $\jvec^{(2)}$, Eq.~(\ref{eq:ddtj_SH}), we have to
express the total real source in terms of the slowly varying
complex amplitudes,
\bea
\vec{S}^{(2)} &=&
\widetilde{\vec{S}}^{(2)}_0 + \left[\widetilde{\vec{S}}^{(2)}_2
\e^{- i 2 \omega_0 t} + \mrm{c.c.}\right]
\nonumber\\
&=& \frac{e\,\epsilon_0}{m_e} \frac{\ompl^2}{\omega_0^2} \nabla
|\widetilde{\vec{E}}^{(1)}|^2 - \frac{e\,\epsilon_0}{m_e} \left\{
\left[ 2 \widetilde{\vec{E}}^{(1)} \left(\nabla \cdot
\widetilde{\vec{E}}^{(1)}\right) +
\frac{1}{2}\frac{\ompl^2}{\omega_0^2} \nabla \left(
\widetilde{\vec{E}}^{(1)} \cdot  \widetilde{\vec{E}}^{(1)} \right)
\right] \e^{-i 2 \omega_0 t} + \mrm{c.c.} \right\}\,.
\label{eq:S_full}
\eea
This result cannot be expressed by the real
linear electric field for all frequencies. But since we are most
interested in the second harmonic generation, we can approximate
the source by \bea \vec{S}^{(2)}\Bigr|_{\mrm{SHG}} &\approx& -
\frac{e\,\epsilon_0}{m_e} \left[ 2 \vec{E}^{(1)} \left(\nabla
\cdot \vec{E}^{(1)}\right) + \frac{1}{2}\frac{\ompl^2}{\omega_0^2}
\nabla |\vec{E}^{(1)}|^2 \right] \,, \label{eq:S_approx} \eea
where $\vec{E}^{(1)}$ is again the full, fast oscillating,
real-valued electric field obtained by the set of
Eqs.~(\ref{eq:ddtB_lin})--(\ref{eq:ddtj_lin}). By inserting the
expansion from Eq.~(\ref{eq:linfreq}) into Eq.~(\ref{eq:S_approx})
it can be easily shown that the second-harmonic contribution of
Eq.~(\ref{eq:S_full}) is exactly reproduced while the
low-frequency contribution of Eq.~(\ref{eq:S_approx}) is different
from that of Eq.~(\ref{eq:S_full}).\footnote{However, in the
spirit of the quasi-monochromatic assumption the spectra around
the second harmonic should be correct. The wrong low-frequency
contribution must be confined to frequencies at most twice the
spectral width of the exciting pulse.}

To numerically solve the $\mathbf{j}_{2}$ equation with the FDTD
approach, Eqs.~(\ref{eq:ddtB_SH})--(\ref{eq:ddtj_SH}) with the
source given by Eq.~(\ref{eq:S_approx}) have to be solved.
Technically, the current is split into three different
contributions according to \bea \frac{\partial
\jvec^{(2)}_A}{\partial t} &=& -\gamma\jvec^{(2)}_A + \frac{e^2
n_0}{m_{e}} \vec{E}^{(2)}\,, \label{eq:j1}
\\
\frac{\partial \jvec^{(2)}_B}{\partial t} &=& -\gamma\jvec^{(2)}_B
- 2 \frac{e\,\epsilon_0}{m_e} \vec{E}^{(1)} \left(\nabla \cdot
\vec{E}^{(1)}\right)\,, \label{eq:j2}
\\
\frac{\partial \jvec^{(2)}_C}{\partial t} &=& -\gamma\jvec^{(2)}_C
- \frac{e\,\epsilon_0}{m_e} \frac{1}{2}\frac{\ompl^2}{\omega_0^2}
\nabla |\vec{E}^{(1)}|^2 \,, \label{eq:j3} \eea where the sum of
$\jvec = \jvec_{A} + \jvec_{B} + \jvec_{C}$ defines the total
current.

\section{Appendix II: Numerical results of third-harmonic generation}

Third-harmonic generation from the samples described in
Ref.\cite{klein2} are numerically simulated, and the obtained
results are listed in Table I. We see that our simulations
reproduce the overall strength of the experimentally observed
third-harmonic signals, and qualitative reproduce the structure
dependent changes. As expected, we find the third-harmonic
strength to be closely related to the localization degree of the
fundamental field inside the metal. In addition, although the
strength of second-harmonic and third-harmonic signals are
comparable \cite{georges}, almost negligible interactions are
observed in our simulations. Second-harmonic generation and
third-harmonic generation from metallic nanoparticles can be
therefore studied separately.

\begin{table}
\caption{Third harmonic strength ($10^{-12}$)}
\begin{ruledtabular}
\begin{tabular}{lcr}
Structure &Experimental results & Numerical results\\
\hline U-shaped (Fig.(1a))      & 3.0  & 1.14\footnotemark[2] \\
       T-shaped (Fig.(1d))  & 0.66& 0.80 \\
       Rectangle (Fig.(1e)) & 0.21 & 0.26 \\
\end{tabular}
\end{ruledtabular}
\footnotetext[1]{The fundamental incidence is $x$-polarized.}
\footnotetext[2]{The difference between experiment and simulation
is possibly due to experimental imperfections in the real
structure, see the Fig.(2a) of Ref.\cite{klein2}.}
\end{table}


\begin{thebibliography}{99}
\bibitem{brown} F. Brown, R. E. Parks, A. M. Sleeper, "Nonlinear Optical Reflection from a Metallic
Boundary", Phys. Rev. Lett. 14, 1029 (1965).
\bibitem{franken} P. A. Franken, A. E. Hill, C. W. Peters, G.
Weinreich, "Generation of Optical Harmonics", Phys. Rev. Lett. 7,
118 (1961).
\bibitem{sipe1} J. E. Sipe and G. I.
Stegeman, {\it Surface Polaritons: Electromagnetic Waves at
Surfaces and Interfaces}, edited by V. M. Agranovich and D. L.
Mills (North-Holland, Amsterdam, 1982).
\bibitem{guidotti} D. Guidotti, T. A. Driscoll, H. J. Gerritsen, "Second harmonic generation in centro-symmetric
semiconductors", Solid State Commun. 46, 337 (1983).
\bibitem{sipe2} J. E. Sipe, D. J. Moss, H. M. van Driel, "Phenomenological theory of optical second- and third-harmonic generation from cubic centrosymmetric
crystals", Phys. Rev. B 35, 1129 (1987).
\bibitem{chen1} C. K. Chen, A. R. B. de Castro, Y. R. Shen, "Surface-Enhanced Second-Harmonic
Generation", Phys. Rev. Lett. 46, 145 (1981).
\bibitem{chen2} C. K. Chen, T. F. Heinz, D. Ricard, Y. R. Shen, "Surface-enhanced second-harmonic generation and Raman
scattering", Phys. Rev. B 27, 1965 (1983).
\bibitem{janz} S. Janz and H. M. van Driel, "Second-harmonic generation from metal
surfaces," International J. Nonlinear Optical Phys. 2, 1(1993).
\bibitem{jha} S. S. Jha, "Theory of optical harmonic generation at a metal surface", Phys. Rev. 140, A2020 (1965).
\bibitem{bloembergen} N. Bloembergen, R. K. Chang, S. S. Jha, C. H.
Lee, "Optical second-harmoni generation in reflection from media
with inverse symmetry", Phys. Rev. 174, 813 (1968).
\bibitem{rudnick} J. Rudnick, E. A. Stern, "Second-harmoni radiation from metal
surfaces", Phys. Rev. B 4, 4272 (1971).
\bibitem{eguiluz} A. Eguiluz, J. J. Quinn, "Hydrodynamic model for surface plasmons in metals and degenerate
semiconductors", Phys. Rev. B 14, 1347 (1976).
\bibitem{sipe3} J. E. Sipe, V. C. Y. So, M. Fukui, G. I. Stegeman,
"Analysis of second-harmonic generation at metal surfaces", Phys.
Rev. B 21, 4389 (1980).
\bibitem{hua} X. M. Hua, J. I. Gersten, "Theory of second-harmonic generation by small metal
spheres", Phys. Rev. B 33, 3756 (1986).
\bibitem{maytorena} J. A. Maytorena, W. Luis Moch\'{a}n, B. S.
Mendoza, "Hydrodynamic model for sum and difference frequency
generation at metal surfaces", Phys. Rev. B 57, 2580 (1998).
\bibitem{sionnest} P. Guyot-Sionnest, W. Chen, Y. R. Shen, "General considerations on optical second-harmonic generation from surfaces and
interfaces", Phys. Rev. B 33, 8254 (1986).
\bibitem{dadap} J. I. Dadap, J. Shan, K. B. Eisenthal, T. F.
Heinz, "Second-Harmonic Rayleigh Scattering from a Sphere of
Centrosymmetric Material", Phys. Rev. Lett. 83, 4045 (1999).
\bibitem{stockman} M. I. Stockman, D. J. Bergman, C. Anceau, S.
Brasselet, J. Zyss, "Enhanced Second-Harmonic Generation by Metal
Surfaces with Nanoscale Roughness: Nanoscale Dephasing,
Depolarization, and Correlations", Phys. Rev. Lett. 92, 057402
(2004).
\bibitem{dadap2} J. I. Dadap, J. Shan, T. F.
Heinz, "Theory of optical second-harmonic generation from a Sphere
of Centrosymmetric Material: small-particle limit", J. Opt. Soc.
Am. B 21, 1328 (2004).
\bibitem{li} K. Li, M. I. Stockman, D. J. Bergman, "Enhanced second harmonic generation in a self-similar chain of metal
nanospheres", Phys. Rev. B 72, 153401 (2005).
\bibitem{cao} L. Cao, N. C. Panoiu, R. M. Osgood, "Surface second-harmonic generation from surface plasmon waves scattered by metallic
nanostructures", Phys. Rev. B 75, 205401 (2007).
\bibitem{bachelier} G. Bachelier, I. Russier-Antoine, E. Benichou, C. Jonin, P. F.
Brevet, "Multipolar second-harmonic generation in noble metal
nanoparticles," J. Opt. Soc. Am. B 25, 955 (2008).
\bibitem{heinz} T. F. Heinz, in {\it Nonlinear surface electromagnetic
phenomena}, edited by H. Ponath and G. Stegeman (Elsevier,
Amsterdam, 1991).
\bibitem{liebsch} A. Liebsch, {\it Electronic Excitations at Metal Surfaces} (Plenum press, 1997).
\bibitem{john} J. B. Pendry, A. J. Holden, D. J. Robbins, W. J.
Stewart, "Magnetism from conductors and enhanced nonlinear
phenomena", IEEE Trans. Microwave Theory Tech. 47, 2075 (1999).
\bibitem{maier} Steven A. Maier, {\it plasmonic: fundamental and
applications} (Springer, 2007).
\bibitem{anceau} C. Anceau, S. Brasselet, J. Zyss, P. Gadenne, "Local second-harmonic generation enhancement on gold nanostructures probed by two-photon
microscopy", Opt. Lett. 28, 713 (2003).
\bibitem{bouhelier} A. Bouhelier, M. Beversluis, A. Hartschuh, and L.
Novotny, "Near-Field Second-Harmonic Generation Induced by Local
Field Enhancement", Phys. Rev. Lett. 90, 013903 (2003).
\bibitem{nahata} A. Nahata, R. A. Linke, T. Ishi, K. Ohashi, "Enhanced nonlinear optical conversion from a periodically nanostructured metal
film", Opt. Lett. 28, 423 (2003).
\bibitem{krause} D. Krause, C. W. Teplin, C. T. Rogers, "Optical surface second harmonic measurements of isotropic thin-film metals: Gold, silver, copper, aluminum, and
tantalum", J. Appl. Phys. 96, 3626 (2004).
\bibitem{nappa} J. Nappa, G. Revillod, I. Russier-Antoine, E.
Benichou, C. Jonin, P. F. Brevet, "Electric dipole origin of the
second harmonic generation of small metallic particles", Phys.
Reb. B 71, 165407 (2005).
\bibitem{neacsu} C. C. Neacsu, G. A. Reider, M. B. Raschke, "Second-harmonic generation from nanoscopic metal tips: Symmetry selection rules for single asymmetric
nanostructures", Phys. Rev. B 71, 201402(R) (2005).
\bibitem{mcMahon} M. D. McMahon, R. Lopez, R. F. Haglund, E. A. Ray, P. H.
Bunton, ``Second-harmonic generation from arrays of symmetric gold
nanoparticles", Phys. Rev. B 73, 041401(R) (2006).
\bibitem{klein} M. W. Klein, C. Enkrich, M. Wegener, S. Linden, "Second-Harmonic Generation from Magnetic Metamaterials", Science 313, 502
(2006).
\bibitem{shan} J. Shan, J. I. Dadap, I. Stiopkin, G. A. Reider, T. F.
Heinz, "Experimental study of optical second-harmonic scattering
from spherical nanoparticles", Phys. Rev. A 73, 023819 (2006).
\bibitem{abe} S. Abe, K. Kajikawa, "Linear and nonlinear optical properties of gold nanospheres immobilized on a metallic
surface", Phys. Rev. B 74, 035416 (2006).
\bibitem{nieuwstadt} J. A. H. van Nieuwstadt, M. Sandtke, R. H. Harmsen, F. B. Segerink, J. C. Prangsma, S. Enoch, L.
Kuipers, "Strong Modification of the Nonlinear Optical Response of
Metallic Subwavelength Hole Arrays", Phys. Rev. Lett. 97, 146102
(2006).
\bibitem{kujala1} S. Kujala, B. K. Canfield, M. Kauranen, Y. Svirko, J. Turunen, "Multipole Interference in the Second-Harmonic Optical
Radiation from Gold Nanoparticles", Phys. Rev. Lett. 98, 167403
(2007).
\bibitem{canfield} B. K. Canfield, H. Husu, J. Laukkanen, B. Bai,
M. Kuittinen, J. Turunen, M. Kauranen, "Local field asymmetry
drives second-harmonic generation in noncentrosymmetric
nanodimers", Nano Lett. 7, 1251 (2007).
\bibitem{klein2} M. W. Klein, M. Wegener, N. Feth, S. Linden,
"Experiments on second- and third-harmonic generation from
magnetic metamaterials", Opt. Express 15, 5238 (2007). Also the
erratum at Opt. Express 16, 8055 (2008).
\bibitem{mcl} M. C. Larciprete, A. Belardini, M. G. Cappeddu, D. de Ceglia, M. Centini, E. Fazio, C. Sibilia, M. J. Bloemer, and M.
Scalora, "Second-harmonic generation from metallodielectric
multilayer photonic-band-gap structures", Phys. Rev. A 77, 013809
(2008).
\bibitem{feth} N. Feth, S. Linden, M. W. Klein, M. Decker, F.
Niesler, Y. Zeng, W. Hoyer, J. Liu, S. W. Koch, J. V. Moloney, and
M. Wegener, "Second-harmonic generation from complementary
split-ring resonators", Opt. Lett. 33, 1975 (2008).
\bibitem{kujala2} S. Kujala, B. K. Canfield, M. Kauranen, Y. Svirko, J. Turunen,
"Multipolar analysis of second-harmonic radiation from gold nanoparticles", Opt. Express, 16, 17196
(2008).
\bibitem{kim} E. Kim, F. Wang, W. Wu, Z. Yu, Y. R. Shen, "Nonlinear optical spectroscopy of photonic metamaterials", Phys. Rev. B
78, 113102 (2008).
\bibitem{maeda} Y. Maeda, T. Iwai, Y. Satake, K. Fujii,
S. Miyatake, D. Miyazaki, G. Mizutani, "Optical second-harmonic
spectroscopy of Au(887) and Au(443) surfaces", Phys. Rev. B 78,
075440 (2008).
\bibitem{husu} H. Husu, B. K. Canfield, J. Laukkanen, B. Bai, M. Kuittinen, J. Turunen, M.
Kauranen, "Chiral coupling in gold nanodimers", Appl. Phys. Lett.
93, 183115 (2008).
\bibitem{rossi} M. Zavelani-Rossi, M. Celebrano, P. Biagioni, D. Polli, M. Finazzi, L. Du\`{o}, G. Cerullo, M. Labardi, M. Allegrini, J. Grand, P.-M.
Adam, "Near-field second-harmonic generation in single gold
nanoparticles", Appl. Phys. Lett. 92, 093119 (2008).
\bibitem{finazzi} M. Finazzi, P. Biagioni, M. Celebrano, L. Du\`{o}, "Selection rules for second-harmonic generation in nanoparticles", Phys. Rev. B 76, 125414 (2007).
\bibitem{walter} W. Hoyer, M. Kira and S. W. Koch, in {\it Classical and quantum optics of semiconductor nanostructures}
(Springer, 2008).
\bibitem{boyd} T. J. M. Boyd, J. J. Sanderson, {\it The physics of plasmas}
(Cambridge, 2003).
\bibitem{jeffrey} J. Freidberg, {\it Plasma Physics and Fusion Energy}
(Cambridge, 2007).
\bibitem{shen} Y. R. Shen, {\it The principles of Nonlinear
Optics} (John Wiley \& Sons, New York, 1984).
\bibitem{robert} R. W. Boyd, {\it Nonlinear
Optics} (Second edition, Academic press, 2003).
\bibitem{taflove} A. Taflove and S. C. Hagness, {\it Computational Electrodynamics: the finite-difference time-domain method} (Second Edition, Artech House, Boston, 2000).
\bibitem{yee} K. Yee, "Numerical solution of initial boundary value problems involving Maxwell's equations in isotropic media", IEEE Trans. Antennas and
Propagat., 14, 302 (1996).
\bibitem{born} M. Born and E. Wolf, {\it Principle of Optics} (Seven Edition, Cambridge, 1999).
\bibitem{palik} E. D. Palik, {\it Handbook of optical constants of solids} (Academic, Orlando, 1985).
\bibitem{berenger} J. P. Berenger, "A perfectly matched layer for the absorption of electromagnetic
waves", J. Computational Physics, 114, 185-200 (1994).
\bibitem{rockstuhl} C. Rockstuhl, F. Lederer, C. Etrich, Th. Zentgraf, J. Kuhl, and H.
Giessen, "On the reinterpretation of resonances in
split-ring-resonators at normal incidence", Opt. Express 14, 8827
(2006).
\bibitem{lazarides} N. Lazarides, M. Eleftheriou, G. P. Tsironis, "Discrete Breathers
in Nonlinear Magnetic Metamaterials", Phys. Rev. Lett. 97, 157406
(2006).
\bibitem{smith} D. R. Smith, W. J. Padilla, D. C. Vier, S. C. Nemat-Nasser, S.
Schultz, "Composite Medium with Simultaneously Negative
Permeability and Permittivity", Phys. Rev. Lett. 84, 4184 (2000).
\bibitem{sa} S. A. Ramakrishna, "Physics of negative refractive index materials", Rep. Prog. Phys. 68, 449 (2005).
\bibitem{yong1} Y. Zeng, X. Chen, W. Lu, "Optical limiting in defective quadratic nonlinear photonic
crystals", J. Appl. Phys. 99, 123107 (2006).
\bibitem{yong2} Y. Zeng, Y. Fu, X. Chen, W. Lu, and H. {\AA}gren, "Highly
efficient generation of entangled photon pair by spontaneous
parametric down-conversion in defective photonic crystal", J. Opt.
Soc. Am. B 24, 1365 (2007).
\bibitem{schaich} W. L. Schaich, "Second harmonic genaration by periodically-structured metal
surfaces", Phys. Rev. B 78, 195416 (2008).
\bibitem{corvi} M. Corvi, W. L. Schaich, "Hydrodynamic-model calculation of second-harmonic generation at a metal
surface", 33, 3688 (1986).
\bibitem{schaich2} W. L. Schaich, A. Liebsch, "Nonretarded hydrodynamic-model calculation of second-harmonic
generation at a metal surface", Phys. Rev. B 37, 6187 (1988).
\bibitem{liebsch2} A. Liebsch, W. L. Schaich, "Second-harmonic generation at simple metal
surfaces", Phys. Rev. B 40, 5401 (1989).
\bibitem{liu} Y. Liu, G. Bartal, D. A. Genov, X. Zhang, "Subwavelength Discrete Solitons in Nonlinear
Metamaterials", Phys. Rev. Lett. 99, 153901 (2007).
\bibitem{feng} S. Feng, K. Halterman, "Parametrically Shielding Electromagnetic Fields by Nonlinear
Metamaterials", Phys. Rev. Lett. 100, 063901 (2008).
\bibitem{zharov} A. A. Zharov, I. V. Shadrivov, Y. S. Kivshar, "Nonlinear Properties of Left-Handed
Metamaterials", Phys. Rev. Lett. 91, 037401 (2003).
\bibitem{scalora} M. Scalora, M. S. Syrchin, N. Akozbek, E. Y.
Poliakov, G. D¡¯Aguanno, N. Mattiucci, M. J. Bloemer, A. M.
Zheltikov, "Generalized Nonlinear Schr\"{o}dinger Equation for
Dispersive Susceptibility and Permeability: Application to
Negative Index Materials", Phys. Rev. Lett. 95, 013902 (2005).
\bibitem{georges} A. T. Georges, "Coherent and incoherent multiple-harmonic generation from metal surfaces", Phys. Rev. A 54, 2412 (1996).
\end{thebibliography}
\end{document}